\documentstyle[12pt]{article}

\begin{document}

\baselineskip=14pt plus 0.2pt minus 0.2pt
\lineskip=14pt plus 0.2pt minus 0.2pt

\begin{flushright}
 LA-UR-93-1721 \\
\end{flushright}

\begin{center}
\Large{\bf
SQUEEZED STATES FOR GENERAL SYSTEMS} \\

\vspace{0.25in}

\large
\bigskip

Michael Martin Nieto$^a$\\
{\it
 $^a$Theoretical Division, Los Alamos National Laboratory\\
University of California\\
Los Alamos, New Mexico 87545, U.S.A. \\}

\vspace{0.25in}

 D. Rodney Truax$^b$\\
{\it $^b$Department of Chemistry, University of Calgary\\
Calgary, Alberta T2N 1N4, Canada\\}

\normalsize

\vspace{0.3in}

{ABSTRACT}

\end{center}
\begin{quotation}
We propose a ladder-operator method for obtaining the squeezed states of
general
 symmetry systems.  It is a generalization of the annihilation-operator
technique
for obtaining the coherent states of symmetry systems.  We connect this method
with
the minimum-uncertainty method for obtaining the squeezed and coherent states
of
general potential systems, and comment on the distinctions between these two
methods and the displacement-operator method.

\vspace{0.25in}


\end{quotation}

\vspace{0.3in}

%\baselineskip=.33in

Coherent states are important in many fields of theoretical and experimental
physics \cite{1,2}.  Similarly, the generalization of coherent states, squeezed
states, has become of more and more interest in recent times \cite{2.5,3}.
This is especially true in the fields of quantum optics \cite{4} and
gravitational
wave
detection \cite{5}.

However, one limitation is that, with the exception we describe below,
essentially all work on squeezed states has concentrated on the harmonic
oscillator
system.  In this letter we describe a generalization of squeezed states to
arbitrary symmetry systems, and its relationship to squeezed states obtained
for
general potentials.

We begin by reviewing coherent states and squeezed states.\\

1) {\it Displacement-Operator Method}. For the harmonic oscillator, coherent
 states are described by the unitary displacement operator acting on the
ground state \cite{6,6.1}:
\begin{equation}
	D(\alpha)|0\rangle  = \exp[\alpha a^{\dagger} - \alpha^* a] |0\rangle
=\exp\left[-\frac{1}{2}|\alpha|^2\right] \sum_{n} \frac{\alpha ^n}{\sqrt{n!}}
|n\rangle
\equiv |\alpha\rangle  .  \label{D}
\end{equation}
The generalization of this method to arbitrary Lie groups has a long history
\cite{1,2,6,9.1}.   One simply applies the displacement operator, which is the
unitary
exponentiation of the factor algebra, on to an extremal state.

As to squeezed states, this method has basically only been applied to  harmonic
oscillator-like systems
 \cite{2.5,3}.
One applies the
SU(1,1) displacement operator onto the coherent state:
\begin{equation}
D(\alpha)S(z)|0\rangle = |(\alpha,z)\rangle,
{}~~~~
S(z) = exp[zK_+ - z^*K_-],  \label{Sop}
\end{equation}
where  $K_+ $,
$K_- $,
and $K_0 $
form an su(1,1) algebra amongst themselves:
\begin{equation}
K_+ = \frac{1}{2}a^{\dagger}a^{\dagger}, \hspace{0.5in}
K_- = \frac{1}{2}aa,  \hspace{0.5in}
K_0 = \frac{1}{2}(a^{\dagger}a + \frac{1}{2}), \label{K}
\end{equation}
\begin{equation}
[K_{0},K_{\pm}] = {\pm}K_{\pm}~,~~~~[K_{+},K_{-}] = -2K_{0}.  \label{KK}
\end{equation}
The ordering of $DS$ vs. $SD$ in Eq. (\ref{Sop}) is unitarily equivalent,
amounting
to a change of parameters.  (Supersymmetric extensions of the above exist
\cite{10}.)\\

2){\it Ladder- (Annihilation-) Operator Method}.  For the harmonic oscillator,
the
coherent states are the eigenstates of the destruction operator:
\begin{equation}
a|\alpha\rangle = \alpha |\alpha\rangle.
\end{equation}
This follows from Eq. (\ref{D}), since
$0 = D(\alpha)a|0\rangle = (a - \alpha)|\alpha \rangle$.
These states are the same as the displacement-operator coherent states.  The
generalization to arbitrary Lie groups is straight forward, and has also been
widely studied \cite{1,2}. \\

3) {\it Minimum-Uncertainty Method}.  This method, which intuitively harks back
to
Schr\"{o}dinger's discovery of the coherent states \cite{12}, has been applied
to
general Hamiltonian potential systems, to obtain both generalized coherent
states
and generalized squeezed states \cite{n1,n2}.  One starts with the classical
problem and transforms it into the ``natural classical variables," $X_c$ and
$P_c$,
which vary as the $\sin$ and the $\cos$ of the classical $\omega t$.  The
Hamiltonian
is therefore of the form $P_c^2 + X_c^2 $.  One then takes these natural
classical
variables and transforms them into ``natural quantum operators."  Since these
are
quantum operators, they have a commutation relation and uncertainty relation:
\begin{equation}
[X,P] = iG, \hspace{0.5in}
(\Delta X)^2(\Delta P)^2 \geq {\frac{1}{4}}\langle G\rangle ^2.
\label{uncert}
\end{equation}
The states
that minimize this uncertainty relation are given by the solutions to the
equation
\begin{equation}
Y\psi_{ss} \equiv
\left(X + \frac{i\langle G\rangle }{2(\Delta P)^2} P\right)\psi_{ss}
=\left(\langle X\rangle +\frac{i\langle G\rangle }{2(\Delta P)^2}\langle
P\rangle
\right)\psi_{ss}.
\end{equation}
Note that of the  four parameters $\langle X\rangle , \langle P\rangle ,
\langle P^2\rangle $, and $\langle G\rangle $, only three are
independent because they satisfy the equality in the uncertainty relation.
Therefore,
\begin{equation}
\left(X + iB P\right)\psi_{ss} = C \psi_{ss}  ,~~~
B = \frac{\Delta X}{\Delta P},  ~~~
 C = \langle X\rangle + i B \langle P\rangle .
\end{equation}
Here $B$ is real and $C$ is complex.  These states, $\psi_{ss}(B,C)$, are the
minimum-uncertainty states for general potentials
\cite{n1,n2}.  Using later parlance, they  are  the squeezed states for general
potentials \cite{3}.  Then $B$ can be adjusted to $B_0$ so that the ground
eigenstate of the potential is a member of the set.  Then these restricted
states,
$\psi_{ss}(B=B_0,C)=\psi_{cs}(B_0,C)$, are the minimum-uncertainty coherent
states
for general  potentials.

It can be intuitively understood that $\psi_{ss}(B,C)$ and $\psi_{ss}(B_0,C)$
are
the
squeezed and coherent states by recalling the situation for the harmonic
oscillator.
The coherent states are the displaced ground state.  The squeezed states are
Gaussians, that have  widths different that that that of the ground state
Gaussian, which are
then displaced.\\

{\it New  Ladder-Operator Method for General Squeezed States}.
General annihilation-operator (or ladder-operator) coherent states are the
eigenstates
of the lowering operator (given a lowest extremal state).
We now propose a generalization to squeezed states, including those for
arbitrary
symmetry
systems:    the general ladder-operator
squeezed states are the eigenstates of a linear combination of the lowering and
raising operators.
(See Comment III, below, concerning previous special cases.)

We will show how the minimum-uncertainty method for obtaining generalized
squeezed states can be used as an intuitive tool to aid in understanding the
ladder-operator method for obtaining generalized squeezed states. We will do
this
with
two specific examples.  Once that is done, the  ladder operator
method can be applied to general symmetry systems, independent of whether they
come from a
Hamiltonian system in the manner of the minimum-uncertainty method above.
Such is our third example.\\

{\it Example I}. First we re-examine the harmonic oscillator, starting  from
the
minimum-uncertainty method.  Here  $X$ and $P$ are obviously $x$ and $p$.  (We
use
dimensionless units.)  Then we have
\begin{equation}
Y = x + s^2 \frac{d}{dx},
\end{equation}
where we have presciently labeled $B$ as $s^2$.  (For the limit to coherent
states,
it turns out that $B=1$.)

Now writing $x$ and $p$ in terms of creation and annihilation operators,
$x =(a + a^{\dagger})/\sqrt{2}, ~ p=(a - a^{\dagger})/(i\sqrt{2})$,
 we find
\begin{equation}
\sqrt{2} \left[
a\left( \frac{1+s^2}{2} \right)
 +  a^{\dagger}
\left( \frac{1-s^2}{2}
\right)
 \right]
\psi_{ss}(s^2,x_0+is^2p_0)
=[x_0+is^2p_0]\psi_{ss}(s^2,x_0+is^2p_0) .
\end{equation}
Therefore, the squeezed states are eigenstates of a linear combination of the
annihilation and creation operators.   Specifically, these states are
\begin{equation}
\psi_{ss}(x) = [\pi s^2]^{-1/4}
\exp\left[-\frac{(x-x_0)^2}{2s^2}+ip_0x\right],   \label{Gauss}
\end{equation}
The relationships to the displacement operator parameters are
$z = re^{i\phi}$, $r = \ln{s}$, $\sqrt{2}\Re{(\alpha)} = x_0$, and
$\sqrt{2}\Im{(\alpha)} = p_0$.
(The phase, $\phi$, is an initial time-displacement.)

We note, with hindsight, that the success of this  method will not be
totally surprising. In many exactly solvable potential systems, the natural
quantum
operators of the minimum-uncertainty method were found to be Hermitian
combinations of the n-dependent raising and lowering operators \cite{n1,n2}.
Here, however, one must generalize to full operators:  $n \rightarrow n(H)$.
Furthermore, in other harmonic-oscillator-like systems, with a Bogoliubov
transformation, this method applies.  (See below.)\\

{\it Example II}.  We demonstrate this  method with the symmetry of the
harmonic
oscillator with centripetal barrier. Previously, the coherent states for this
particular example were found with
the minimum-uncertainty method, but not the squeezed states \cite{n2}.
Therefore,
it
is an ideal system since, at the end, we can connect to the coherent states
obtained
from the minimum-uncertainty method.

This system contains an su(1,1) algebra \cite{truax}. Its elements are
\begin{equation}
L_{\pm} =\frac{1}{4\nu}\frac{d^2}{dz^2}
\mp \frac{1}{2} z \frac{d}{dz} \mp \frac{1}{4} + \frac{\nu}{4} z^2
- \frac{\nu}{4 z^2}  ,
\end{equation}
\begin{equation}
L_0 = \frac{H}{4\nu} + \frac{\nu}{2}  , ~~~~
H = -  \frac{d^2}{dz^2}
+ \nu^2 \left(\frac{1}{z} - z\right)^2 .  \label{L0}
\end{equation}
In terms of the $X$ and $P$ minimum-uncertainty operators \cite{n2}, we find
\begin{equation}
X = \frac{L_- + L_+}{\nu} = z^2 - \left(1+\frac{H}{2\nu^2}\right), ~~~
P = \frac{2(L_- -L_+)}{i} = \frac{1}{i}\left[2z\frac{d}{dz} + 1\right].
\end{equation}
Therefore, the squeezed states for this system are formed by the
solution to the equation
\begin{equation}
0=\left[y\frac{d^2}{dy^2} + \left(\frac{1}{2} + 2\nu By\right)\frac{d}{dy}
+\frac{1}{4}\left(y-\frac{\nu^2}{y}+2B\nu\right)-\frac{\nu C}{2}\right]
\psi_{ss},
\end{equation}
where we have changed variables to $y=\nu z^2$.
The squeezed state solutions to this
equation are
\begin{equation}
\psi_{ss} = N\exp[-y(\nu B + \gamma)]
\left[y^{\lambda + \frac{1}{2}}\right]
\Phi\left(\left[\frac{\nu C}{4\gamma}+\frac{1}{2}(\lambda+\frac{3}{2})
\right],~
\left[\lambda+\frac{3}{2}\right];~2\gamma y\right),
\end{equation}
where $\Phi(a,b;c)$ is the confluent hypergeometric function
$ \sum_{n=0}^{\infty} \frac{(a)_n c^n}{(b)_n~n!}$,
$\gamma =\sqrt{\nu^2 B^2 - \frac{1}{4}} $, and
$ \lambda(\lambda +1) = \nu^2$.
In the limit where $B
\rightarrow 1/(2\nu)$, these become the coherent states given in  Ref.
\cite{n2},
\begin{equation}
\psi_{cs} = \left[\frac{2\nu^{1/2}e^{-\nu \Re(C)}}{I_{\lambda + 1/2}(\nu |C|)}
\right]^{1/2}
e^{-y/2} y^{1/4}I_{\lambda +1/2}\left((2\nu Cy)^{1/2}\right),
\end{equation}
where $I$ is the modified Bessel function. \\

{\it Example III}.  We now consider a symmetry system which does not have as
its
origin a  Hamiltonian system.  We consider the su(1,1) symmetry of Eqs.
(\ref{K}, \ref{KK}).  Our ladder-operator squeezed states are thus the
solutions to
\begin{equation}
\left[\left(\frac{1+s}{2}\right) aa
+\left(\frac{1-s}{2}\right) a^{\dagger} a^{\dagger} \right]\psi_{ss}
=\beta^2 \psi_{ss} .   \label{Kss}
\end{equation}
where the analogue of $B$ is $s$ and the role of $C$ is taken by $\beta^2$.
Using the differential representations of the ladder operators,  Eq.
(\ref{Kss})
can be written in the form
\begin{equation}
 \left[\frac{d^2}{dy^2} +2ys\frac{d}{dy} +y^2 +(s-2\beta^2)\right]
\psi_{ss} = 0.
\end{equation}

Observe that  the ladder operators raise and lower the number states by two
units.
Therefore, there will be two solutions to this equation, one containing only
even
number states and one containing only odd number states.  We will designate
these as
$\psi_{Ess}$ and $\psi_{Oss}$.  These solutions are
\begin{equation}
\psi_{Ess}=N_E\exp{\left[-\frac{-y^2}{2}(s+\sqrt{s^2-1})\right]}
\Phi\left(\left[\frac{1}{4}+\frac{\beta^2}{2\sqrt{s^2-1}}\right],~
\frac{1}{2};~y^2\sqrt{s^2-1}\right),
\end{equation}
\begin{equation}
\psi_{Oss}=N_O ~y\exp{\left[-\frac{-y^2}{2}(s+\sqrt{s^2-1})\right]}
\Phi\left(\left[\frac{3}{4}+\frac{\beta^2}{2\sqrt{s^2-1}}\right],~
\frac{3}{2};~y^2\sqrt{s^2-1}\right).
\end{equation}
In the limit $s\rightarrow 1$, these become the even and odd coherent states:
\begin{equation}
\psi_{Ecs}=
\left[\frac{e^{-\beta^2}}{\pi^{1/2}\cosh{|\beta|^2}}\right]^{1/2}
\exp\left[-\frac{1}{2}y^2\right]\cosh(\sqrt{2}\beta y),
\end{equation}
\begin{equation}
\psi_{Ocs}=
\left[\frac{e^{-\beta^2}}{\pi^{1/2}\sinh{|\beta|^2}}\right]^{1/2}
\exp\left[-\frac{1}{2}y^2\right]\sinh(\sqrt{2}\beta y).
\end{equation}
Using generating formulae, these can be written in the number-state basis as
\begin{equation}
\psi_{Ecs}\ = [\cosh{|\beta|^2}]^{-1/2}
\sum_{n=0}^{\infty}\frac{\beta^{2n}}{\sqrt{(2n)!}}|2n\rangle ,
\end{equation}
\begin{equation}
\psi_{Ocs}\ = [\sinh{|\beta|^2}]^{-1/2}
\sum_{n=0}^{\infty}\frac{\beta^{2n+1}}{\sqrt{(2n+1)!}}|2n+1\rangle .
\end{equation}
Up to the normalization, these are the ``even and odd coherent states"
previously
found in Ref. \cite{mankoEO}.
Although this system did not come from a Hamiltonian, one could have
used a minimum-uncertainty principle to obtain the same states by starting with
the commutation relation
$[K_{+},K_{-}] = -2K_{0}$.
However, one does not obtain the same coherent states from
the displacement-operator method.  Those coherent states, defined by
$S(z)|0\rangle$,   are  the squeezed-vacuum Gaussian of Eq. (\ref{Gauss}) with
$x_0 = p_0 = 0$.  \\

{\it Comment I}.  The above discussion brings us to the displacement-operator
method.
Although it is the natural method for defining coherent states for Lie
algebras,
there is
as yet no well-known general extension of this method to define general
displacement-operator squeezed
states.  This last has been touched upon in discussions \cite{sq1} about
higher-order generalizations of the ``squeeze operator," $S(z)$.  In
particular,
although harmonic-oscillator like systems admit squeeze operators (or
Bogoliubov
transformations) connecting  the  displacement-operator and ladder-operator
methods
\cite{ass,Agar}, the appropriate generalization of these squeeze operators have
not
been found.  Therefore, for now, the ladder-operator method is generally
connected
only to the minimum-uncertainty method.\\

{\it Comment II}.  In this vein, for finite-dimensional representations, such
as for angular momentum coherent states, the ladder-operator method does not
allow a
solution for coherent states, although the displacement-operator method does
\cite{ass}.  Contrariwise, for squeezed states, we observe that the opposite is
true.\\

{\it Comment III}.  The above three examples have all been cases where
$A_- = (A_+)^{^{\dagger}}$.  Sometimes that is not the case, as in certain
potential systems
whose eigenenergies are not equally spaced \cite{n1,n2}.  Then, as in Eq.
(\ref{L0}), one should use the operator form for ``$n$": $A_n \rightarrow
A_{n(H)}$, to connect to the minimum-uncertainty method.  In these cases, the
ladder-operator coherent and sueezed states can be different than, though
related
to, their minimum-uncertainty counterparts.

\vspace{1.0in}

\noindent  Email:  $^a$mmn@pion.lanl.gov, $^b$truax@acs.ucalgary.ca

\end{document}